\documentclass[aps,pre,twocolumn,floatfix,nofootinbib,showpacs]{revtex4}
\usepackage[dvips]{graphicx}
\usepackage{epsf,amssymb,amsmath,mathptmx}
\begin{document}
\title{Bidding process in online auctions and winning strategy: 
rate equation approach}
\author{I.~Yang and B. Kahng}
\affiliation{School of Physics and Center for Theoretical Physics,
Seoul National University, Seoul 151-747, Korea}
\date{\today}
\begin{abstract}
Online auctions have expanded rapidly over the last decade and have 
become a fascinating new type of business or commercial transaction 
in this digital era. Here we introduce a master equation for the
bidding process that takes place in online auctions. 
We find that the number of distinct bidders who bid $k$ times, 
called the $k$-frequent bidder, up to the $t$-th bidding progresses 
as $n_k(t)\sim tk^{-2.4}$. 
The successfully transmitted bidding rate by the $k$-frequent
bidder is obtained as $q_k(t) \sim k^{-1.4}$, independent of
$t$ for large $t$. This theoretical prediction is in agreement with
empirical data. These results imply that bidding at the
last moment is a rational and effective strategy to win in an eBay
auction.
\end{abstract}
\pacs{89.75.-k, 89.75.Da, 89.65.-s, 89.65.Gh}
\maketitle
Electronic commerce (e-commerce) refers to any type of
business or commercial transaction that involves information
transfer across the Internet. As a formation of e-commerce, 
the online auction, i.e., the auction via the Internet~\cite{heck}, 
has expanded rapidly over the last decade and has become a 
fascinating new type of business or commercial transaction in 
this digital era. Online auction technology has several benefits 
compared with traditional auctions. Traditional auctions require 
the simultaneous participation of all bidders or agents at the 
same location; these limitations do not exist in online auction 
systems. Owing to this convenience, ``eBay.com," the largest 
online auction site, boasts over 40 million registered consumers 
and has experienced rapid revenue growth in recent years.

Interestingly, the activities arising in online auctions generated
by individual agents proceed in a self-organized 
manner~\cite{mantegna,bouchard, stanley,challet,pennock,hulst}. For
example, the total number of bids placed in a single item or
category and the bid frequency submitted by each agent follow power-law
distributions~\cite{yang}. These power-law behaviors~
\cite{simon,zhang,albert} are rooted in the fact that an agent who
makes frequent bids up to a certain time is more likely to bid
in the next time interval. This pattern is theoretically analogous 
to the process that is often referred to as preferential attachment, 
which is responsible for the emergence of scaling in complex
networks~\cite{ba}. This is reminiscent of the mechanism of 
generating the Zipf law~\cite{zhang,pareto}. The accumulated data of
a detailed bidding process enable us to quantitatively characterize 
the dynamic process. In this paper, we describe a master equation 
for the bidding process. 
The master-equation approach is useful to capture the dynamics 
of the online bidding process because it takes into account of the effect 
of openness and the non-equilibrium nature of the auction. 
This model is in contrast to the existing equilibrium approach~\cite{lb,kauf} 
in which there is a fixed number of bidders. The equilibrium approach 
is relevant to traditional auctions; however, it is unrealistic 
to apply this approach to Internet auctions. The power-law 
behavior of the bidding frequency submitted by individual agents 
can be reproduced from the master equation. 
Moreover, we consider the probability of an agent 
who has bidden $k$ times, called the $k$-frequent 
bidder, becoming the final winner. We conclude that the winner is 
likely to be the one who bids at the last moment but who placed 
infrequent bids in the past.

Our study is based on empirical data collected from two
different sources~\cite{yang}. The first dataset was downloaded
from the web, http://www.eBay.com, and is composed of all the auctions
that closed in a single day. The data include 264,073 auctioned items,
grouped into 194 subcategories. The dataset allows us
to identify 384,058 distinct agents via their unique user IDs. To
verify the validity of our findings in different markets and time
spans, the second dataset was accumulated over a period of one year from
eBay's Korean partner, auction.co.kr. The dataset comprised 215,852 agents
that bid on 287,018 articles in 355 lowest categories.

An auction is a public sale in which property or items of merchandise
are sold to the bidder who proposes the highest price. Typically,
most online auction companies adopt the approach of English auction, in
which an article or item is initially offered at a low price that is
progressively raised until a transaction is made. Both ``eBay.com"
and ``auction.co.kr" adopt this rule and many bidders submit multiple 
bids in the course of the auction. An agent is not allowed to place 
two or more bids in direct succession. It is important to notice 
that the eBay auction has a fixed end time: It typically ends a 
week after the auction begins, at the same time of day to 
the second. The winner is the latest agent to bid within this period. 
In such an auction that has a fixed deadline, bidding that takes 
place very close to the deadline does not give other bidders sufficient 
time to respond.
In this case, a sniper--the last moment bidder--might win the
auction, while the bid that follows has a substantial probability 
of not being transmitted successfully. While such a bidding pattern is
well known empirically, no quantitative analysis has been
performed on it as yet. In this study we analyze this issue through 
the rate equation approach.

To characterize the dynamic process, we first introduce several
quantities for each item or article as follows:
\begin{itemize}
\item[(i)] When a bid is successfully transmitted, time $t$ 
increases by one. 
\item[(ii)] Terminal time $T$ is the time at which an 
auction ends. Thus, the index of bids runs from $i=1$ to $T$.
\item[(iii)] $N(t)$ is the number of distinct bidders who 
successfully bid at least once up to time $t$. 
Thus, the index of bidders (or agent) runs from $i=1$ to $N(t)$. 
\item[(iv)] $k_i(t)$ is the number of successful
bids transmitted by an agent $i$ up to time $t$.
\item[(v)] $n_k(t)$ is the number of bidders with 
frequency $k$ up to time $t$.
\end{itemize}
From the above, we obtain the relations
\begin{equation}
N(t)=\sum_k n_k(t)
\end{equation} and
\begin{equation} t=\sum_{k} k n_{k}(t)
\end{equation}
for any time $t$ including the terminal time $T$.

\begin{figure}[t]
\centerline{\epsfxsize=8cm \epsfbox{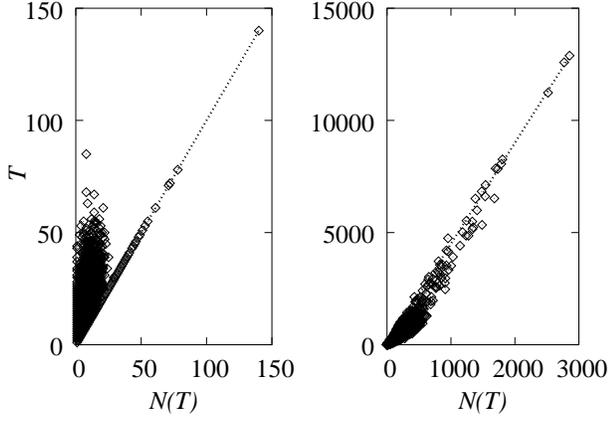}}
\caption{Plot of $T$ versus $N(T)$ for the eBay.com (a) and the
Korean auction (b). The dotted line has a slope of 1 both in (a) 
and 4.5 in (b).} \label{fig:K_N_each}
\end{figure}

\begin{figure}
\centerline{\epsfxsize=8cm \epsfbox{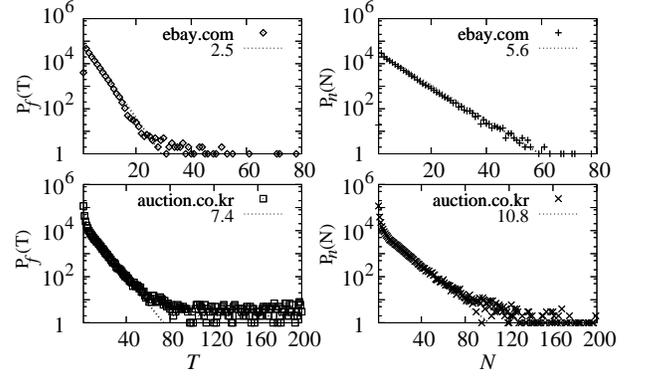}} 
\caption{Plot of $P_f(T)$ versus $T$ in (a) and (c), and $P_n(N)$ 
versus $N$ in (b) and (d) for the eBay  (a) and (b) and Korean 
auction (c) and (d)
in semi-logarithmic scale. The dotted lines have slopes of $2.5$ in
(a), $5.6$ in (b), $7.4$ in (c), and $10.8$ in (d).} \label{fig2}
\end{figure}

It is numerically found that $T$ is linearly proportional to $N(T)$,
that is, $T \propto N(T)$. The average value of the
proportional coefficient $a$ for different items or articles listed 
in eBay is estimated to be $a \approx 1$ when the total number
of bidders $N(T)$ exceeds $20$. However, when the number of
bidders is lower, the proportional coefficient is very large, as
shown in Fig.~\ref{fig:K_N_each}. For the Korean auction, $a
\approx 4.5$, regardless of the number of bidders. On the other
hand, the bidding frequencies and the number of bidders for each
article are not uniform. Their distributions, denoted as $P_f (T)$
and $P_n (N)$, respectively, follow the exponential functions $P_f (T)\sim
\exp(-T/T_c)$ and $P_n(N)\sim \exp(-N/N_c)$, respectively, where
$T_c\approx 7.4$ and 10.8 for the eBay and Korean auctions,
respectively, and $N_c\approx 2.5$ and 5.6 for the eBay and 
Korean auction, respectively (Fig.\ref{fig2}).

\begin{figure}
\centerline{\epsfxsize=8cm \epsfbox{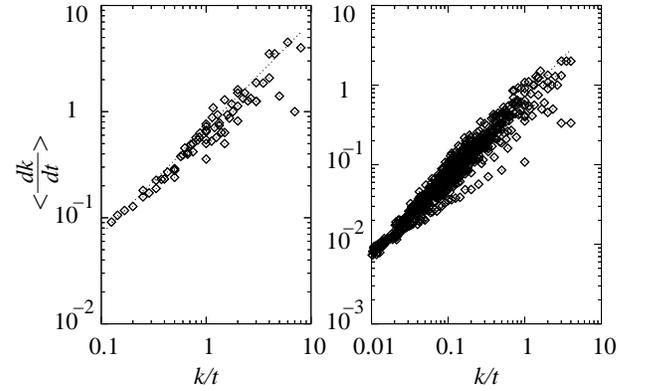}}
\caption{Plot of $\langle dk/dt \rangle$ versus $k/t$ for the eBay (a)
and for the Korean auction (b). The dotted lines obtained 
by the least square fit in the range [0.1:1] (a) and 
[0.01:1] (b), respectively, fit to the formula, $\approx 0.7k/t$.}
\label{fig:w_k_t}
\end{figure}

We introduce the master equation for the bidding process as
\begin{equation}
n_k(t+1)-n_k(t)=w_{k-1}(t)n_{k-1}(t)-w_k(t)n_k(t)+\delta_{k,1}u_t,
\label{discrete}
\end{equation}
where $w_k(t)$ is the transition probability that a bidder, who
has bid $k-1$ times up to time $t-2$, bids at time $t$. 
In this case, the total bid frequency of that agent up to time $t$ 
becomes $k$.
Note that a bidder is not allowed to bid successively. In the
master equation, we presume that the bidding pattern is similar
over different items when $N(T)$ is sufficiently large. Then,
${w_k}(t)$ may be written as ${w_k}(t) \approx \langle dk/dt
\rangle$ on average over different items. Empirically, we find
that
\begin{equation}
{w_k}(t)\approx \langle dk/dt \rangle \approx bk/t, \label{kernel}
\end{equation}
where $b$ is estimated to be $b\approx 0.7$ for both the eBay and
Korean auctions (Fig.~\ref{fig:w_k_t}). The fact that $w_k
\propto k$ is reminiscent of the preferential attachment rule in
the growing model of the complex network~\cite{ba}. $u_t$ is the
probability that a new bidder makes a bid at time $t$. Using the
property that $\sum_k n_k (t)=N(t)$, we obtain 
\begin{equation}
u_t=N(t+1)-N(t).
\end{equation}

Next we then change the discrete equation, Eq.~(\ref{discrete}), to
a continuous equation as follows:
\begin{equation}
\frac {\partial n_k(t)}{\partial t}=-\frac {\partial}{\partial
k}\big({w_k}(t)n_k(t)\big)+\delta_{k,1} u_t, \label{continuous}
\end{equation}
which can be rewritten as
\begin{equation}
\frac {\partial n_k(t)}{\partial t}=-\frac{b}{t}\frac
{\partial}{\partial k}\big(k n_k(t)\big)+\delta_{k,1} u_t.
\label{continuous2}
\end{equation}
When $k > 1 $, we use the method of separation of variables,
$n_k(t)=I(k)T(t)$, thus obtaining 
\begin{equation}
\frac{\partial}{\partial k}(k I(k))+\ell I(k)=0, \label{eq_K}
\end{equation}
where $\ell$ is a constant of separation, and
\begin{equation}
\frac{\partial T(t)}{\partial t}= \frac {b\ell}{t}T(t).
\end{equation}
Thus, we obtain 
\begin{equation}
n_k(t)\sim t^{b\ell} k^{-(1+\ell)}.
\end{equation}
When $k=1$,
\begin{eqnarray}
\frac{\partial n_1(t)}{\partial t}=-\frac{b}{t}n_1(t)
+u_t.
\label{eq:n_1}
\end{eqnarray}

\begin{figure}
\centerline{\epsfxsize=8cm \epsfbox{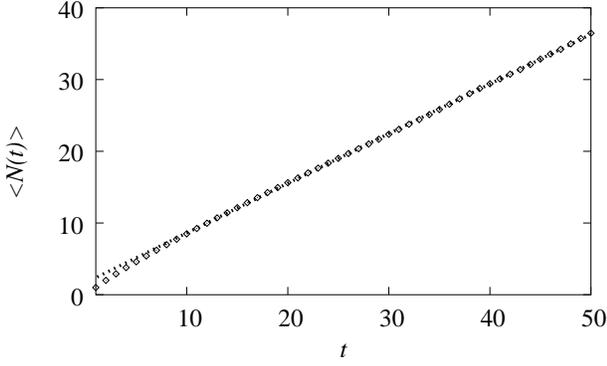}}
\caption{Plot of $\langle N(t) \rangle$ versus $t$, on average
over different items for the eBay data. The straight line has a 
slope of $0.7$ obtained from the least square fit.
 } \label{fig:N_t_fig}
\end{figure}

\begin{figure}[b]
\centerline{\epsfxsize=8cm \epsfbox{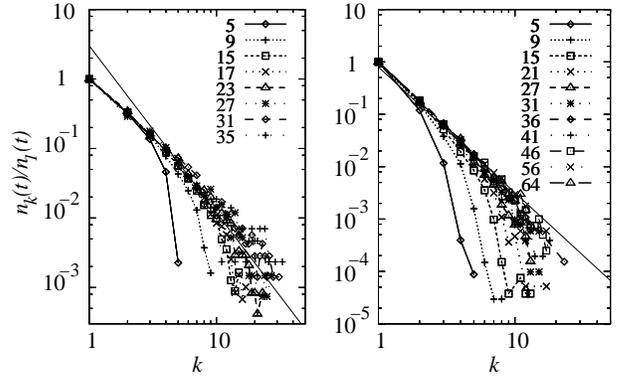}}
\caption{Plot of $n_k$ versus $k$ for the eBay auction 
at ebay.com (a) and for the Korean auction at
auction.co.kr (b) for various terminal times $T$. The solid lines
have a slope of -2.4 drawn for guidance.}
\label{fig:N_DNST}
\end{figure}

Next from the fact that $N=\sum_k n_k$, we obtain
\begin{eqnarray*}
\frac{\partial N}{\partial t} &=&
\sum_{k > 1}\frac{\partial n_k}{\partial t}+\frac{\partial n_1}{\partial t}\\
&=&\sum_{k>1}-\frac{b}{t}\frac{\partial}{\partial k}\Big(kn_k\Big)
-\frac{b}{t}n_1+\frac{\partial N}{\partial t}\\
&=& \frac{b\ell}{t}(N-n_1)-\frac{b}{t}n_1+\frac{\partial N}{\partial t}.
\end{eqnarray*}
Therefore, we obtain $N(t)=(1+1/\ell)n_1(t)$ and $n_1(t)\sim
t^{b\ell}$ by using Eq.~(\ref{eq:n_1}). Note that $N(t) < t$, and
the linear relationship holds asymptotically. The linear relationship 
breaks down for small $t$. From the empirical data,
Fig.~\ref{fig:N_t_fig}, we find that $\ell b\approx 1$. Since
$b\approx 0.7$ in Fig.~\ref{fig:w_k_t}, we obtain $\ell
\approx 1/b\approx 1.4$. Therefore,
\begin{equation}
n_k(t)\sim t k^{-2.4} \label{n_k} \end{equation} 
for large $t$, which fits reasonably with the numerical data 
shown in Fig.\ref{fig:N_DNST}.

In eBay auctions, the winner is the last bidder in the bidding
sequence. Now, we trace the bidding activity of the winner in the
bidding sequence in order to find the winning strategy. To
proceed, let me define $q_{k}(t+1)$ as the probability that a
bidder, who has bid $k-1$ times up to time $t-1$, bids at time
$t+1$ successfully. Note that a bidder is not allowed to bid 
successively. In this case, $q_k(T)$ is nothing but the probability that a
$k$-frequent bidder becomes the final winner. The probability
$q_k(t+1)$ satisfies the relation,
\begin{eqnarray}
q_{k}(t+1)&=&
(1-u_{t+1})\sum_{j=1}^{N(t)}q_{j}(t)\frac{(k-1)(n_{k-1}(t)-
\delta_{j,k-1})}{t-j}\nonumber \\
&+&\delta_{k,1}u_{t+1}\label{eq:win}
\end{eqnarray}
with the boundary conditions $q_{1}(1)=1$ and $q_{1}(2)=1$. The
first term on the right hand side of Eq.~(\ref{eq:win}) is
composed of three factors: (i) $1-u_{t+1}$ is the
probability that one of the existing bidders bids successfully at
time $t+1$, (ii) $q_j(t)$ means that bidding at
time $t$ is carried out by the $j$-frequent bidder, and (iii) the last
factor is derived from the bidding rate, Eq.~(\ref{kernel}), where the
contribution by the bidder at time $t$ is excluded because he/she
is not allowed to bid at time $t+1$. The second term represents
the addition of a new bidder at time $t$.

\begin{figure}
\centerline{\epsfxsize=8cm \epsfbox{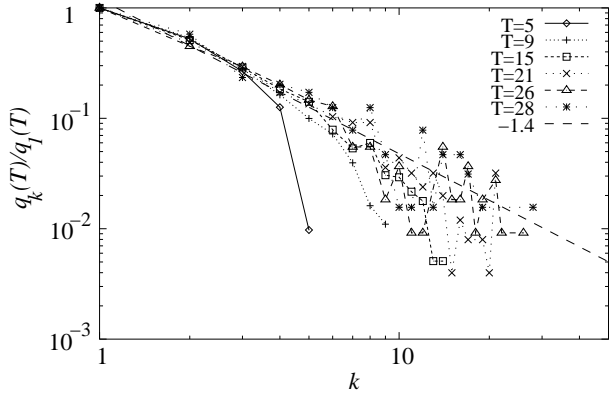}}
\caption{Plot of the relative winning probability $q_k(T)/q_1(T)$
of the $k$-frequent bidder to that of the one-frequent bidder at
the last moment versus frequency $k$. 
The dotted line has a slope of -1.4 drawn for guidance.} 
\label{fig:winning}
\end{figure}

The rate equation, Eq.~(\ref{eq:win}), can be solved recursively.
To proceed, we simplify Eq.~(\ref{eq:win}) by assuming that
$n_{k-1}(t)$ is significantly larger than $\delta_{j,k-1}$, which is
relevant when the number of bidders is large. Then,
\begin{widetext}
\begin{eqnarray}
q_k(t+1)& \approx & (1-u_{t+1})
\sum_{i=1}^{N(t)}q_i(t)\frac{(k-1)n_{k-1}(t)}
{t-i}+\delta_{k,1}u_{t+1}\nonumber \\
&=& (k-1)n_{k-1}(t)\prod_{\tau=2}^{t} (1-u_{\tau+1})
\Big[ \sum_{i=1}^{\tau-1} \frac {(i-1)n_{i-1}(\tau)}{(\tau-i)} \Big] q_{1}(2) \\
&+&(1-u_{t+1})(k-1) n_{k-1}(t)\sum_{\tau=3}^{t}  \frac{
u_{\tau}}{\tau-1}\prod_{\tau'=\tau+1}^{t}(1-u_{\tau'}) \Big[
\sum_{i=1}^{\tau^{\prime}-1} \frac
{(i-1)n_{i-1}(\tau^{\prime})}{(\tau^{\prime}-i)}\Big]+
u_{t+1}\delta_{k,1}\nonumber.
\end{eqnarray}
\end{widetext}
Since $1-u_t\approx 0.3 < 1$, $q_k(t)$ is obtained to be
\begin{equation}
q_k(t)\approx
(1-u_{t-1})\frac{(k-1)n_{k-1}(t-1)}{t-2}+\delta_{k,1}u_t
\end{equation}
within the leading order. Considering that $n_k(t)\sim
tk^{-2.4}$ in Eq.~(\ref{n_k}) and $u_t$ is constant, we obtain
$q_k(t)\sim (t-1)k^{-1.4}/(t-2)$ for large $k$ and $t$,
with a weak dependence on $t$. Thus, the winning probability
by the $k$-frequent bidder is simply given as 
\begin{equation}
q_k(T)\sim k^{-1.4} \end{equation}
in the limit $t\to \infty$. This result is confirmed by the
empirical data in Fig.~\ref{fig:winning}.

Our analysis explicitly shows that the winning strategy is to bid 
at the last moment as the first attempt rather than
incremental bidding from the start. This result is consistent
with the empirical finding by Roth and Ockenfels~\cite{roth} in
eBay. According to them, the bidders who have won the most items tend 
to wait till the last minute to submit bids, albeit there is some 
probability of bids not being successfully transmitted. 
As evidence, they studied 240 eBay auctions and found that 89 
bids were submitted in the last minute and
29 in the last ten seconds. Our result supports these empirical
results.

In conclusion, we have analyzed the statistical properties of
emerging patterns created by a large number of agents based on the
empirical data collected from eBay.com and auction.co.kr. 
The number of bidders and the winning probability
decay in power laws as $n_k(t)\sim tk^{-2.4}$ and $q_k(t)\sim
k^{-1.4}$, respectively, with bid frequency $k$, which
has been confirmed by empirical data.\\

This work is supported by the KRF Grant No. R14-2002-059-01000-0
in the ABRL program funded by the Korean government MOEHRD and 
the CNS research fellowship in SNU (BK).

\end{document}